# Label-free Imaging of Single-Biomolecule Structure and Interaction by Stimulated Raman Photothermal Encoded Scattering

Pin-Tian Lyu[1,2], Yifan Zhu[1,2], Qing Xia[1,2], Guangrui Ding[1,2], Arvind Pillai[3,4,5], Xinru Wang[4,5,6], Jianpeng Ao[1,2], Haonan Lin[1,2,7], Lulu Jiang[8,9], David Baker[4,5,10], Ji-Xin Cheng[1,2]*

[1]Department of Electrical and Computer Engineering, Boston University; Boston, MA 02215, USA.

[2]Photonics Center, Boston University; Boston, MA 02215, USA.

[3]Stowers Institute for Medical Research; Kansas City, MO 64110, USA.

[4]Department of Biochemistry, University of Washington; Seattle, WA 98195, USA.

[5]Institute for Protein Design, University of Washington; Seattle, WA 98195, USA.

[6]Center for Drug Discovery, Departments of Pharmaceutical Sciences and Chemical Biology, Northeastern University; Boston, MA 02115, USA.

[7]The Wallace H. Coulter Department of Biomedical Engineering, Georgia Institute of Technology and Emory University; Atlanta, GA 30332, USA.

[8]Department of Neuroscience, University of Virginia School of Medicine; Charlottesville, VA 22908, USA.

[9]Center for Brain Immunology and Glia (BIG), University of Virginia School of Medicine; Charlottesville, VA 22908, USA.

[10]Howard Hughes Medical Institute, University of Washington; Seattle, WA 98195, USA.

*Corresponding author. Email: jxcheng@bu.edu

**Abstract**

Current single molecule methods either rely on fluorescence or lack chemical information. Here we report stimulated Raman photothermal encoded scattering (SRPSCAT) microscopy for quantitative bond-selective imaging of single-biomolecule structures and interactions in native environments. In this approach, scattering of the target molecule is modulated by the deposited energy from stimulated Raman gain and loss processes, thereby encoding vibrational spectroscopic information. Leveraging single-molecule sensitivity of interferometric scattering, SRPSCAT can map single proteins with chemical specificity, determine their mass, and distinguish protein secondary structures based on their Raman fingerprints. Furthermore, single protein binding kinetics are quantified and the conformational dynamics of single de novo designed allosteric proteins are observed. Together, these results highlight the potential of SRPSCAT for label-free structural, functional and dynamic analysis at the single-molecule level.

## Introduction

Optical detection of single molecules has fundamentally transformed our understanding of chemical and biological processes, opening new avenues for gene expression mechanism (*1*), super-resolution imaging (*2*, *3*), and biomolecule mass quantification (*4*, *5*). Despite tremendous advances in single-molecule fluorescence microscopy capable of characterizing interactions and mechanisms with high specificity (*6–8*), the labels provide limited information about the structure and function of the host molecule and are subject to photobleaching. As a label-free approach, light scattering is highly sensitive and universally applicable for detecting and characterizing molecules. Based on elastic scattering, interferometric scattering (iSCAT) microscopy (*4*, *5*, *9*) and plasmonic/evanescent scattering microscopy (*10*, *11*) have succeeded in measuring single-protein mass and interactions, yet lacking chemical specificity and structural information.

To gain chemical and structural information of single molecules, surface-enhanced Raman scattering (SERS) has enabled Raman spectroscopy of single molecules adsorbed on nanostructured surfaces (*12*, *13*), but the non-uniformity of 'hot spots' and strict reliance on close contact to the nanostructures limit its applications in biological systems. Derived from SERS, surface-enhanced coherent Raman scattering techniques (*14–17*) achieve high-speed vibrational imaging of single molecules, nevertheless, face the same restrictions as SERS. A recent approach called stimulated Raman excited fluorescence (SREF) encodes vibrational resonance into fluorescence excitation and achieves far-field Raman imaging of single molecules (*18*). However, SREF is only applicable to dye molecules with the electronic pre-resonance scheme. Thus far, no method provides bond-selective chemical information at the single-molecule level in a label-free and far-field manner.

Here we report stimulated Raman photothermal encoded scattering (SRPSCAT) microscopy that achieves quantitative single-molecule vibrational imaging in their native environments. This method harnesses the photothermal effect induced by stimulated Raman gain and loss processes (*19*, *20*) to encode vibrational spectroscopic information into iSCAT, enabling label-free single-molecule spectroscopic imaging.

## Results

### SRPSCAT principle, setup and performance

In a stimulated Raman scattering (SRS) process, the stimulated Raman gain and loss deposit energy ($\omega_p - \omega_s$) to the target molecule through vibrational excitation (**Fig. 1A**). The subsequent relaxation of vibrational energy via nonradiative decay heats up the molecules and surrounding environments, known as the stimulated Raman photothermal (SRP) effect (*20*). On the other hand, elastic scattering (**Fig. 1A**) offers superb sensitivity and has been used in visible photothermal imaging of single dye molecules (*21*, *22*). As a common-path interferometric measurement of scattering, iSCAT detects the scattered field of the molecule interfered with the reference field from the interface (**Fig. 1B**). Because of its single-molecule sensitivity and broad applicability, iSCAT has enabled imaging of single-molecule interactions (*23*, *24*) and single ion channel activity on cells (*25*), and ion (*26*) and carrier (*27*, *28*) transport in materials. Here, we encode vibrational information into iSCAT via the SRP effect, termed SRPSCAT. In this method, SRS excitation produces heat and causes the temperature rise ($\Delta T$) of the target molecule, followed by thermal expansion ($\Delta r$) and refractive index change ($\Delta n$), which modulates the scattering field thus the iSCAT signal (**Fig. 1C**). Theoretical modeling shows that the modulation depth of SRPSCAT for a single immunoglobulin M (IgM) molecule can reach the order of $10^{-5}$ under the experimental conditions, where the transient temperature increase is calculated to be ~7.76 K. In comparison, the modulation depth of SRS is at $10^{-7}$ level under the same conditions, which is 2 orders of magnitude lower than that of SRPSCAT (fig. S1 and Supplementary Text).

To fulfill the SRPSCAT concept, we have built an SRPSCAT microscope as shown in **Fig. 1D**. Briefly, the pump and Stokes pulses are intensity-modulated and combined. A probe beam is aligned collinearly with the SRS beams. The back-scattered light by the sample and the reflected light at the sample-substrate interface are collected by the objective, spectrally filtered, and detected by a photodiode followed by an amplifier. The output signal is demodulated by a lock-in amplifier and recorded with a data acquisition card. More details can be found in Methods and fig. S2. Hyperspectral imaging is achieved by tuning the Raman shift ($\omega_p - \omega_s$) in spectral focusing mode. Several strategies are applied to extract the weak photothermal modulation signals from iSCAT photons. First, 1-MHz modulation rejects low-frequency noise and enables shot-noise limited detection. Second, substrates with high refractive index ($n = 1.7$) are used to ensure sufficient reflected photons for demodulation. Third, a self-supervised deep learning framework, self-permutation Noise2Noise denoiser (SPEND), is employed to recover signal-to-noise ratio (SNR) from system complexity (*29*).

We characterized the performance of SRPSCAT using standard polystyrene (PS) nanoparticles. **Fig. 1E** shows an SRPSCAT hyperspectral image of single PS particles of 29 nm in diameter at 3055 $cm^{-1}$ corresponding to the aromatic C–H stretching mode. After SPEND processing that improves SNR by 8-fold (*29*) , the image exhibits much lower noise level than the raw image, allowing the weak signal from the nanoparticles to be clearly resolved. As shown in **Fig. 1F**, the acquired SRPSCAT spectrum from a single PS particle shows the Raman signature at 3055 $cm^{-1}$, demonstrating the high spectral fidelity, while SRS imaging of the same 29 nm PS particles shows no contrast with identical average laser power (**Fig. 1G**). The SRS spectrum only shows the cross-phase modulation background (**Fig. 1H**). In such case, the signal cannot be rescued by SPEND (fig. S3).

To validate the imaging mechanism of SRPSCAT, we imaged PS nanoparticles with varied sizes. The SRPSCAT images of PS nanoparticles with diameters of 29 nm, 50 nm and 75 nm at 3055 $cm^{-1}$ are shown in **Fig. 1I**. The SRPSCAT spectrum of each particle is plotted in **Fig. 1J**, exhibiting consistent spectral shape. It is clearly shown in the image and spectra that SRPSCAT intensity increases with increasing particle sizes. For each particle size, we counted the intensity from individual particles and constructed an intensity histogram, which follows a Gaussian distribution (fig. S4). The mean intensity versus nanoparticle diameter is plotted in logarithmic scale, revealing that the image intensity follows a power law of $d^3$ (**Fig. 1K**). This indicates that the SRPSCAT signal is dominated by the interference term rather than the pure scattering term, which decreases rapidly with particle size ($d^6$). Though the detection of single nanoparticles can be achieved by other label-free scattering approaches, SRPSCAT offers valuable chemical selectivity in the spectral domain. For example, SRPSCAT could differentiate individual 50-nm poly(methyl methacrylate) (PMMA) and PS particles by their Raman signatures, showing the potential in characterizing single nanoparticles with high sensitivity and chemical specificity (fig. S5).

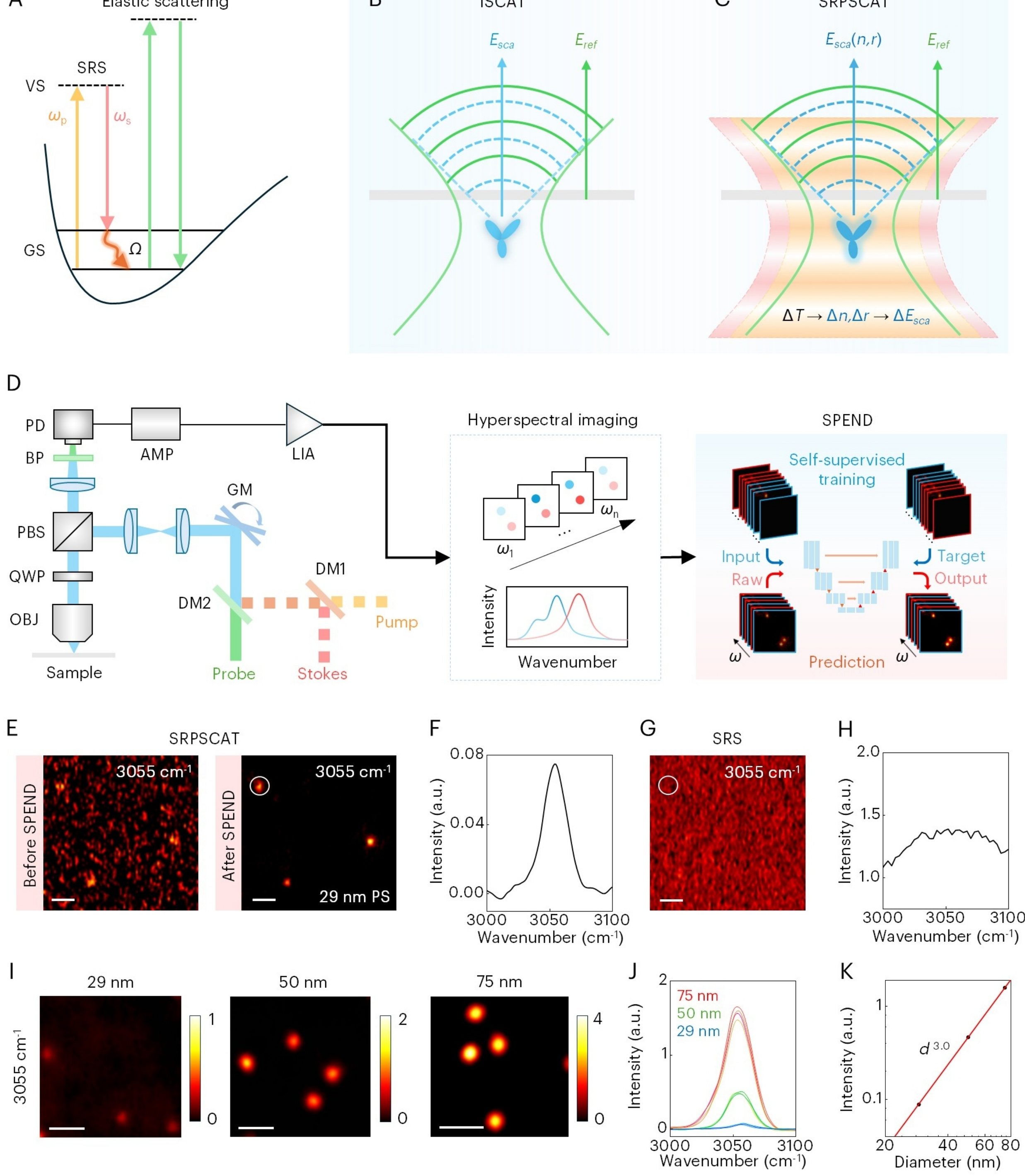

**Fig. 1. Principle, setup and performance of SRPSCAT.**
(**A**) Energy diagram of SRS and elastic scattering. GS, ground state; VS, virtual state.
(**B**) Schematic of wavefronts of laser illumination (solid lines) and sample radiation (dashed lines) in scanning mode iSCAT.
(**C**) Schematic of SRPSCAT with pump (orange) and Stokes (red) illumination and iSCAT probe.
(**D**) SRPSCAT setup, hyperspectral image acquisition and deep learning-based image denoising process by self-permutation Noise2Noise denoiser (SPEND). DM, dichroic mirror; GM, galvo scanning mirror; QWP, quarter-wave plate; OBJ, objective; BP, bandpass spectral filter; PD, photodiode; AMP, amplifier; LIA, lock-in amplifier.
(**E**) Raw and denoised SRPSCAT images of single 29 nm PS particles.
(**F**) SRPSCAT spectra of a single PS particle circled in (E).
(**G**) Raw SRS image at the same field of view of (E) with the same average laser power.
(**H**) SRS spectrum at the circled area in (G) showing cross-phase modulation.
(**I**) SRPSCAT imaging of PS particles with different diameters.
(**J**) SRPSCAT spectra of single PS particles in (I).
(**K**) SRPSCAT intensity versus PS particle diameter. Scale bars, 1 μm.

### Quantitative mass imaging of single proteins with chemical specificity

Based on the approximation that the specific volume of amino acids and the refractive indices of proteins are constant, a linear relation between molecular mass and iSCAT contrast at the single-molecule level has been observed (*4*, *5*, *30*). The iSCAT signal is, however, not chemically sensitive. To demonstrate the capability of SRPSCAT for single-protein mass imaging with chemical information, we first studied human IgM with the molecular mass of 950 kDa. The IgM solution was incubated and then dried on the glass surface to form dispersed single proteins or clusters as characterized by atomic force microscopy (AFM) shown in **Fig. 2A**. The line profiles of three spots marked in Fig. 2A are shown in **Fig. 2B**. The measured height of ~4 nm and horizontal diameter of ~38 nm for single IgM matches the reported value (*31*). We also observed double- and triple-stacked IgM clusters with multiple heights. Next, we conducted iSCAT and SRPSCAT measurements of the dried IgM samples. Though iSCAT shows image contrasts of individual spots (**Fig. 2C**), the SRPSCAT image at the same field of view identifies IgM specifically at 2939 $cm^{-1}$ corresponding to the $CH_3$ stretching mode in immunoglobulin (*32*) (**Fig. 2D**). Notably, not all spots in the iSCAT image give the SRPSCAT signal (marked by the white circle in Fig. 2D), indicating the existence of non-protein substances, such as dust or salts from the medium, on the glass surface. These substances that exhibit similar contrasts to the target molecules challenge elastic scattering methods. In contrast, SRPSCAT not only measures signal intensity as elastic scattering does, but also provides chemical specificity to identify proteins at a spatial resolution of ~230 nm (fig. S6).

Then, we statistically investigated the SRPSCAT spectra and intensity of individual proteins and clusters. As shown in **Fig. 2E**, the SRPSCAT spectra can be classified into different groups according to their separated intensity, exhibiting quantized distributions and integer multiple relationships. This indicates the observation of single, double and triple IgM, respectively. It is noteworthy that the spectral widths are narrower than those of the protein solution (*33*). One possible reason is that the solution measurements show the result of ensemble average while single-molecule experiments reveal the heterogeneity. In addition, we found a decrease in the ratio of 2880 $cm^{-1}$ to 2939 $cm^{-1}$ peaks in these dried samples compared with proteins in solution because denatured proteins exposed to water molecules have lower relative intensity at 2880 $cm^{-1}$ (*34*, *35*). To confirm the observation of single proteins, we counted the SRPSCAT intensity at 2939 $cm^{-1}$ from individual spots and constructed an intensity histogram (**Fig. 2F**). By fitting the histogram with three isolated Gaussian functions, the mean intensity of each distribution was extracted, from which the linear relation between the intensity and the mass of single, double and triple IgM was established (**Fig. 2G**). The good linear fitting further supports the SRPSCAT imaging of single IgM and its clusters, showing the feasibility of quantitative mass imaging with SRPSCAT.

To validate the linear relation between molecular mass and SRPSCAT intensity, we carried out SRPSCAT measurements for five different proteins spanning 66.5 kDa to 950 kDa (**Fig. 2H** & fig. S7). The proteins immobilized on the glass surface were immersed in buffer solution during measurements. We found that the increasing protein mass leads to an increase in SRPSCAT intensity. More importantly, we observed the $CH_3$ peak shift between immunoglobulins (2939 $cm^{-1}$), e.g., human IgM, human immunoglobulin A (IgA) and human immunoglobulin G (IgG), and other proteins (2930 $cm^{-1}$), e.g., human transferrin and bovine serum albumin (BSA). This shift is correlated with the protein secondary structure. The rich β-sheet structures in immunoglobulin brings more restriction to the movement of side chains than the relatively flexible α-helix structures in albumin and transferrin, leading to the shift of the C–H band to the high wavenumber side (*34*). The β-sheet structure in IgM was measured in the Raman fingerprint window with SRPSCAT (fig. S8). The amide I band at 1671 $cm^{-1}$ shows good agreement with the peak assignment of β-sheet structures in literature (*36*). Additionally, compared with the denatured proteins in dry environments, we observed a normal ratio of 2880 $cm^{-1}$ to 2930 $cm^{-1}$/2939 $cm^{-1}$ peaks (Fig. 2H) in these native proteins in solution as reported (*33*).

To statistically study the SRPSCAT intensity of different proteins, we monitored single proteins as they diffused from solution to nonspecifically bind to the glass surface. The dynamic binding process allows us to rapidly track and count the individual molecule binding events for each protein and construct corresponding intensity histogram. The SRPSCAT intensity histogram of the individual molecules follows a Gaussian distribution (**Fig. 2I**). There is a small second peak, which may be attributed to the formation of dimers or two molecules binding to the nearby surface simultaneously with distance smaller than the diffraction limit.

To determine the relationship between protein mass and SRPSCAT intensity, we obtained the mean intensity of these proteins by fitting the histograms with Gaussian functions and plotted the intensity versus protein mass (**Fig. 2J**). The linear relation further supports the results of single protein imaging and validates the capability of SRPSCAT for quantitative mass imaging of single biomolecules. The mass accuracy, i.e. the deviation between the measured and sequence mass, was determined to be $<5$ kDa, resulting in an average error of 2%. The detection sensitivity has reached single BSA proteins of 66.5 kDa in the C–H region. Overall, the performance of SRPSCAT for mass measurements is comparable to standard iSCAT mass photometry (*5*). Meanwhile, we show that SRPSCAT provides valuable insight into biomolecular structures and their interactions with environments, which are key to their functions.

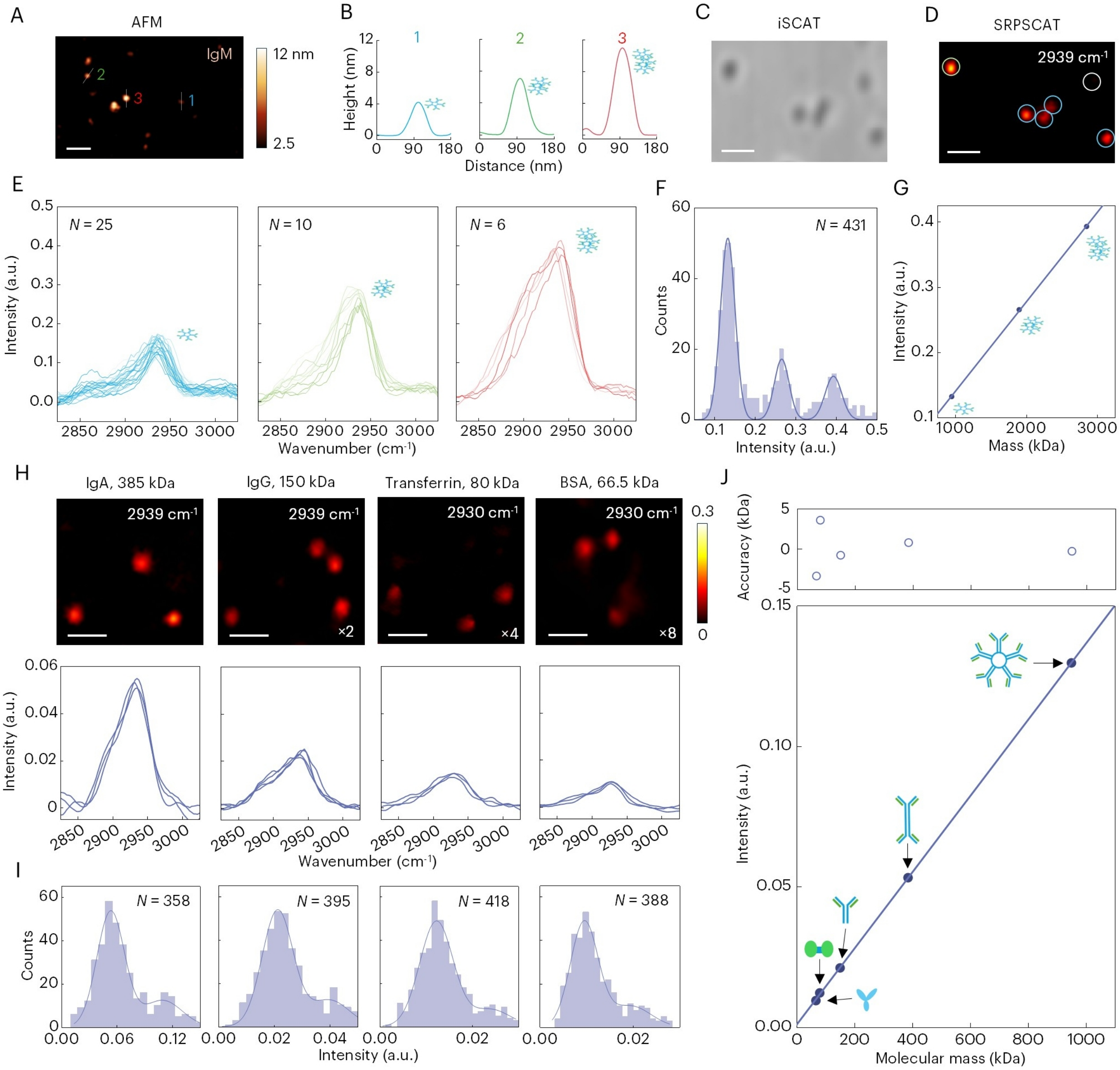

**Fig. 2. Quantitative mass imaging of single proteins with SRPSCAT.**

(**A** and **B**) AFM image (A) and line profiles (B) of IgM on coverslip. 1: Single IgM, 2: Double IgM, 3: Triple IgM. Scale bar: 200 nm.

(**C** and **D**) iSCAT image (C) and SRPSCAT image (D) of IgM at the same field of view. Blue circles: Single IgM, Green circle: Double IgM, White circle: Non-protein substance. Scale bars, 1 μm.

(**E**) SRPSCAT spectra of single (blue lines), double (green lines) and triple (red lines) IgM.

(**F**) SRPSCAT intensity histogram of IgM at 2939 $cm^{-1}$ fitted with three Gaussian functions.

(**G**) SRPSCAT intensity versus mass of single, double and triple IgM with linear fitting. The intensity is obtained from the mean value of the corresponding Gaussian distribution.

(**H**) SRPSCAT images of single IgA, IgG, transferrin and BSA at their peak intensity and the corresponding SRPSCAT spectra. The color bars are scaled according to the factor displayed in each image. Scale bars, 1 μm.

(**I**) SRPSCAT intensity histograms of the corresponding proteins. In each histogram, the main Gaussian peak is from single proteins and a small second peak is attributed to protein dimers or two protein binding to the surface with distance smaller than the diffraction limit.
(**J**) SRPSCAT intensity versus molecular mass. Proteins plotted in the order of mass: BSA, transferrin, IgG, IgA and IgM. The intensity is obtained from the mean value of the corresponding histogram. Mass accuracy (upper panel) is given as the difference between the sequence mass and the value suggested by the linear fit.

**Elucidation of single protein secondary structure**

Beyond molecular mass, proteins form certain folded structures to drive their physiological functions. Elucidating protein structures is fundamental to deciphering physiological cellular mechanisms and their pathological dysregulation in disease. To this end, vibrational microscopy techniques, including infrared and Raman spectroscopic imaging, have been widely employed to study protein structures via characteristic chemical bond vibrations (*36–39*). However, imaging the secondary structure of a single protein in its native state remains a major challenge.

We first performed SRPSCAT imaging of apoferritin and fibronectin by targeting the C=O vibrational band sensitive to protein secondary structure. Apoferritin is the iron-removed shell of ferritin with the mass of 440−480 kDa and is rich in α-helix (**Fig. 3A**), while fibronectin is a glycoprotein of the extracellular matrix with almost identical mass (440−500 kDa) to apoferritin but contains a high β-sheet content (**Fig. 3D**). The SRPSCAT image and spectra of individual apoferritin show the typical Raman band of α-helix at 1656 $cm^{-1}$ (**Fig. 3, B and C**). The Raman signature of β-sheet in individual fibronectin at 1671 $cm^{-1}$ can also be clearly resolved with SRPSCAT (**Fig. 3, E and F**). To further confirm single-protein imaging with chemical selectivity, we measured a mixed sample of apoferritin and fibronectin. It is difficult to differentiate these two proteins in the iSCAT image because their masses are very close (**Fig. 3G**). In contrast, SRPSCAT can distinguish them by providing Raman signatures of secondary structures with spectral unmixing (**Fig. 3H**). The unmixing map was obtained by a pixel-wise least absolute shrinkage and selection operator (see details in Methods) (*33*). Notably, although the spectral shift in the C–H region can be resolved for these two proteins because of their distinct structures (fig. S9), the fingerprint window provides richer structural information with narrower bandwidth compared to the broad C–H band.

With the capability for single-protein structural elucidation, we further applied SRPSCAT to study structural transitions during the aggregation of intrinsically disordered proteins. In particular, we focused on α-synuclein (αS), whose misfolding and aggregation are central to the pathogenesis of Parkinson's disease and related synucleinopathies (*40*). The transition of αS from a largely disordered monomer into β-sheet–rich oligomers and fibrils has been shown to mediate its neurotoxicity, particularly the soluble oligomers and pre-formed fibrils (**Fig. 3I**) (*41*, *42*). Resolving the secondary structure of individual heterogeneous oligomers is critical to understanding their aggregation mechanism and cellular toxicity (*43*), however, remains challenging to existing methods such as cryogenic electron microscopy (*44*). Here, we investigated the secondary structure of individual αS oligomers by SRPSCAT (**Fig. 3J**). The slight peak shift and varied peak width associated with different peak intensity indicates that the secondary structure of different oligomeric species is highly heterogeneous. The αS fibrils, in contrast, show a narrower and homogeneous spectral feature dominated by highly ordered β-sheet (**Fig. 3K**). Furthermore, a peak fitting approach based on the Raman signature of αS was employed to quantitatively analyze the contribution of secondary structural elements in αS monomers, oligomers and fibrils (*45*, *46*). As shown in **Fig. 3L**, the Raman bands are deconvolved into α-helix (~1656 $cm^{-1}$), β-sheet (~1671 $cm^{-1}$), extended structures (~1680 $cm^{-1}$) and the ring modes of aromatic residues (~1615 $cm^{-1}$). The αS monomers with random structures show a broadband Raman spectrum. The Raman band narrows dramatically and the β-sheet component increases during the aggregation, indicating the formation of more ordered aggregates. An increase of β-sheet from 39% in oligomers to 59% in fibrils was observed with a concomitant decrease of α-helix from 24% to 13%. These changes demonstrate the conformational transition during αS aggregation. Together, with the capability to distinguish monomeric, oligomeric and fibrillar structures, SRPSCAT provides single-molecule insights into aggregation mechanisms, structural heterogeneity, and disease-relevant conformational changes, enabling new opportunities for fundamental studies and therapeutic intervention.

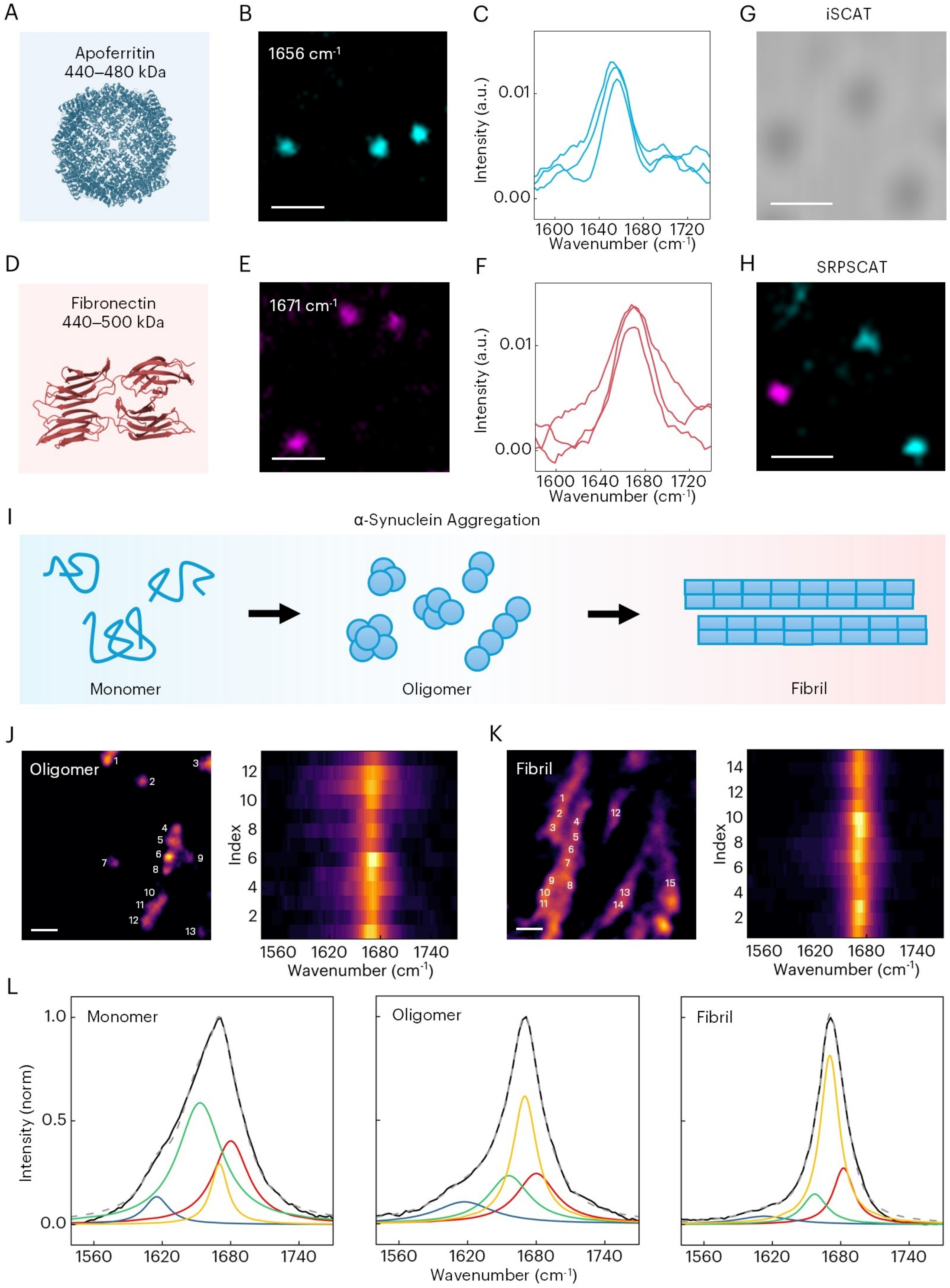

**Fig. 3. Elucidating protein secondary structure with SRPSCAT.**
(**A**) Schematic of α-helix dominated chain in apoferritin. Image adapted from the RCSB Protein Data Bank (PDB ID: 2W0O) (*47*).
(**B** and **C**) SRPSCAT image (B) and spectra (C) of individual apoferritin in fingerprint window.
(**D**) Schematic of β-sheet dominated fragment in fibronectin. Image adapted from the RCSB Protein Data Bank (PDB ID: 6MFA) (*47*).
(**E** and **F**) SRPSCAT image (E) and spectra (F) of individual fibronectin in fingerprint window.

(**G** and **H**) iSCAT image (G) and SRPSCAT spectral unmixing map (H) of a mixed apoferriagtin and fibronectin sample at the same field of view. Scale bars, 1 μm.
(**I**) Schematic of αS aggregation pathway.
(**J** and **K**) SRPSCAT images and spectra of individual αS oligomers (J) and fibrils (K). The index correlates each spectrum with the image. Scale bars, 1 μm.
(**L**) Averaged SRPSCAT spectra of αS monomer solution, oligomers and fibrils with peak fitting (raw spectra in black, ring modes in blue, α-helix in green, β-sheet in orange, extended structures in red, sum spectra in grey).

**Quantifying single protein binding kinetics**

After having demonstrated the capabilities of SRPSCAT for quantifying molecular mass and imaging protein structures, we sought to monitor biomolecular dynamics, such as protein-protein interactions and binding kinetics, at the single-molecule level. This is essential for understanding molecular mechanisms and functions, thereby facilitating drug discovery.

As a demonstration, we investigated the interaction between IgA and its antibody, anti-IgA, at the single-molecule level. We first immobilized single anti-IgA on glass surface, which will be used as a sensor surface for binding (**Fig. 4A**). The SRPSCAT image and corresponding spectra show that this anti-IgA with an immunoglobulin structure has a $CH_3$ stretching peak at 2939 $cm^{-1}$ (**Fig. 4, B and C**). At this wavenumber, we counted the intensity of individual molecules and constructed an intensity histogram (**Fig. 4D**). The mean intensity further confirmed that this anti-IgA is an isotype of IgG according to the mass calibration curve in Fig. 2J.

To show the capability of monitoring dynamic binding events of single molecules, we tracked floating IgA that nonspecifically binds to the glass-solution interface at 2939 $cm^{-1}$ (**Fig. 4E**). The binding process of single IgA was observed as bright spots appearing one at a time on the surface. Movie S1 reveals an entire binding process and **Fig. 4F** shows a few snapshots. After repeating the measurements, we counted the intensity of hundreds of spots and constructed a histogram, showing a major peak from single IgA according to the same mass calibration curve (**Fig. 4G**). We note that the minor peaks show at the position with double intensity of the mean value in both Fig. 4D and Fig. 4G are likely contributed by protein dimers or two molecules simultaneously falling within a diffraction-limited area.

To identify the specific binding events and quantify protein binding kinetics at the single-molecule level, we applied IgA solution to the anti-IgA coated sensor surface (**Fig. 4H**). The immobilized anti-IgA are visualized as individual spots with certain intensity and the binding and unbinding events of single IgA can be tracked by the intensity change of the spot at 2939 $cm^{-1}$ in real time. This enables monitoring of the heterogeneity of protein behaviors, due to different conformation, orientation and location on the sensor surface. To illustrate this point, **Fig. 4I to K** and Movies S2 to S4 show three different behaviors of binding between individual anti-IgA and IgA molecules, including an IgA molecule (1) binds tightly to an anti-IgA, (2) binds to an anti-IgA temporally, then leaves the surface and (3) binds and unbinds rapidly. From the temporal intensity profile of IgA molecule 3 (fig. S10), the distributions of residence times of the bound and unbound states are obtained as shown in **Fig. 4L and M**. Fitting of the distributions with the first-order binding kinetics model (see Supplementary Text for details) determines the association ($k_{on}$) and dissociation ($k_{off}$) rate constants of this single IgA molecule, which are $1.3\times10^9$ $M^{-1}$ $s^{-1}$ and 7.8 $s^{-1}$, respectively. From $k_{on}$ and $k_{off}$, the equilibrium dissociation constant ($K_D = k_{off}/k_{on}$) is determined to be 6.1 nM. These values are in good agreement with the previously reported results (*10*). Furthermore, analyzing the binding kinetics for different molecules reveals the heterogeneity of protein binding (fig. S11). These results demonstrate that SRPSCAT can recognize protein-protein interactions and quantify protein binding kinetics at the single-molecule level. It is noteworthy that SRPSCAT allows visualization of each molecule during their interactions so neither differential image processing nor blocking nonspecific binding sites is needed.

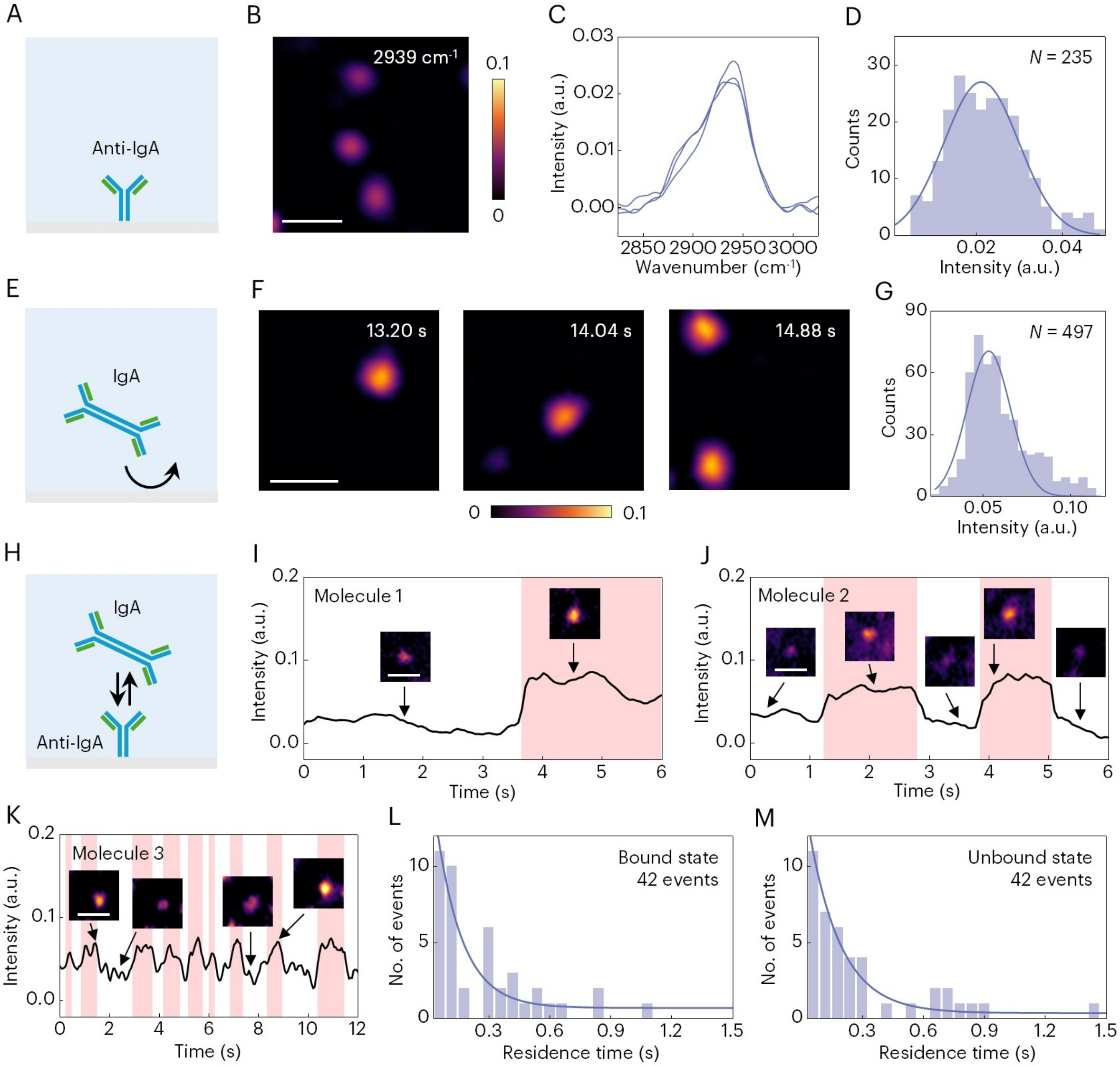

**Fig. 4. Observing single protein binding kinetics with SRPSCAT.**
(**A**) Schematic of anti-IgA immobilized on glass surface.
(**B** and **C**) SRPSCAT image (B) and spectra (C) of single anti-IgA.
(**D**) SRPSCAT intensity histogram of anti-IgA fitted with a Gaussian function.
(**E**) Schematic of floating IgA that lands on glass surface then leaves.
(**F**) SRPSCAT snapshots of individual IgA landing events.
(**G**) SRPSCAT intensity histogram of IgA fitted with a Gaussian function.
(**H**) Schematic of binding of IgA to anti-IgA immobilized on glass surface.
(**I** to **K**) Examples of different binding behaviors of IgA molecule 1 (I), 2 (J) and 3 (K).

(**L** and **M**) Residence time distribution of bound (L) and unbound (M) states for molecule 3 fitted with exponential functions, from which $k_{on}$, $k_{off}$ and $K_D$ of molecule 3 are determined to be $1.3\times10^9$ $M^{-1}$ $s^{-1}$, 7.8 $s^{-1}$ and 6.1 nM, respectively. Scale bars, 1 μm.

### Monitoring conformational dynamics of de novo designed allosteric proteins

Protein conformation is intrinsic to its activity and functionality. The functions of many proteins are regulated through allostery (*48*), whereby effector binding induces the conformational transformation that controls cell metabolism and signaling. Unveiling real-time conformational dynamics of allosteric proteins at the single-molecule level will facilitate the understanding of their allosteric communications and dynamic cellular processes. However, progress in this area has been constrained by the limited availability of accessible single-molecule techniques, as well as the high cost and technical complexity of existing methodologies such as solution nuclear magnetic resonance, single-molecule fluorescence resonance energy transfer, etc.

To address this challenge, we investigated the capability of SRPSCAT to study protein conformational state changes in real time. As a proof of concept, we applied it to a well-defined system, a de novo designed allosteric protein oligomer, sr312, which switches between two distinct oligomeric conformations in response to effector binding (*49*). In absence of effectors, sr312 is a compact triangular trimetric ring shape with a small pore (X3 state, **Fig. 5A**). Upon the binding of the effector peptide, this architecture remodels into a square-like tetrameric ring, accompanied by a marked expansion of the internal pore diameter and an increase in protein mass (Y4 state, **Fig. 5B**). We first measured the protein samples tethered to the glass-solution interface with SRPSCAT. The SRPSCAT images and spectra of individual sr312 molecules in X3 state and Y4 state show the characteristic peaks of proteins in native environment at 2930 $cm^{-1}$ (**Fig. 5, C to F**). To confirm the imaging of homogenous oligomeric protein complexes , we counted the intensity of individual spots and constructed intensity histograms from dynamic binding measurements, thereby quantifying the mass of sr312 in both states based on the mass calibration curve (**Fig. 5, G and H**, and fig. S13). The mass of sr312 in X3 state and Y4 state are measured to be 142.2 kDa and 317.0 kDa, respectively. These results are in consistency with the mass measured by mass photometry and amino acid sequence identity (*49*), demonstrating the observation of proteins of single oligomeric state. Moreover, we measured another de novo designed protein, sr508, which remains in the same oligomeric state on addition of effectors but undergoes allosteric toggling in conformation. SRPSCAT measurements of mass confirmed that its oligomeric state remains unchanged. The results further support single-protein imaging and mass quantification (fig. S12).

To reveal the conformational dynamics at the single-protein level, we tethered sr312 X3 molecules to a modified glass surface and then added effector peptide solution immediately before measurements. The mechanism of conformational change is illustrated in **Fig. 5I**. The conformation of the triangular X3 state will change from linear to V-shaped when the effector binds to the switch, leading to the protein assembly into square-like Y4 state. The associated increase in protein mass allows us to monitor conformational dynamics by tracking the SRPSCAT intensity at 2930 $cm^{-1}$ in real time. The temporal intensity profiles of three individual sr312 molecules show one-step intensity increase, consistent with a switch from X3 state to Y4 state (**Fig. 5, J to L**). We increased the speed to 30 ms per frame, enabling capturing transient conformational change at video rate by reducing the imaging area (fig. S14, fig. S15 and Movie S5). These results show that there is an instantaneous transition from X3 state to Y4 state within 30 ms, which is consistent with the characteristic time scale of allostery in natural proteins (*50*–*53*). In comparison, the area without immobilized molecules typically shows no contrast in the time window (fig. S16). The temporal intensity fluctuation around the mean value is likely attributed to the bulk heating of the medium and the motion of molecules in the medium. Collectively, we demonstrate that SRPSCAT enables high-speed screening of conformational switching in computationally designed proteins and provides a new approach for studying protein conformational dynamics.

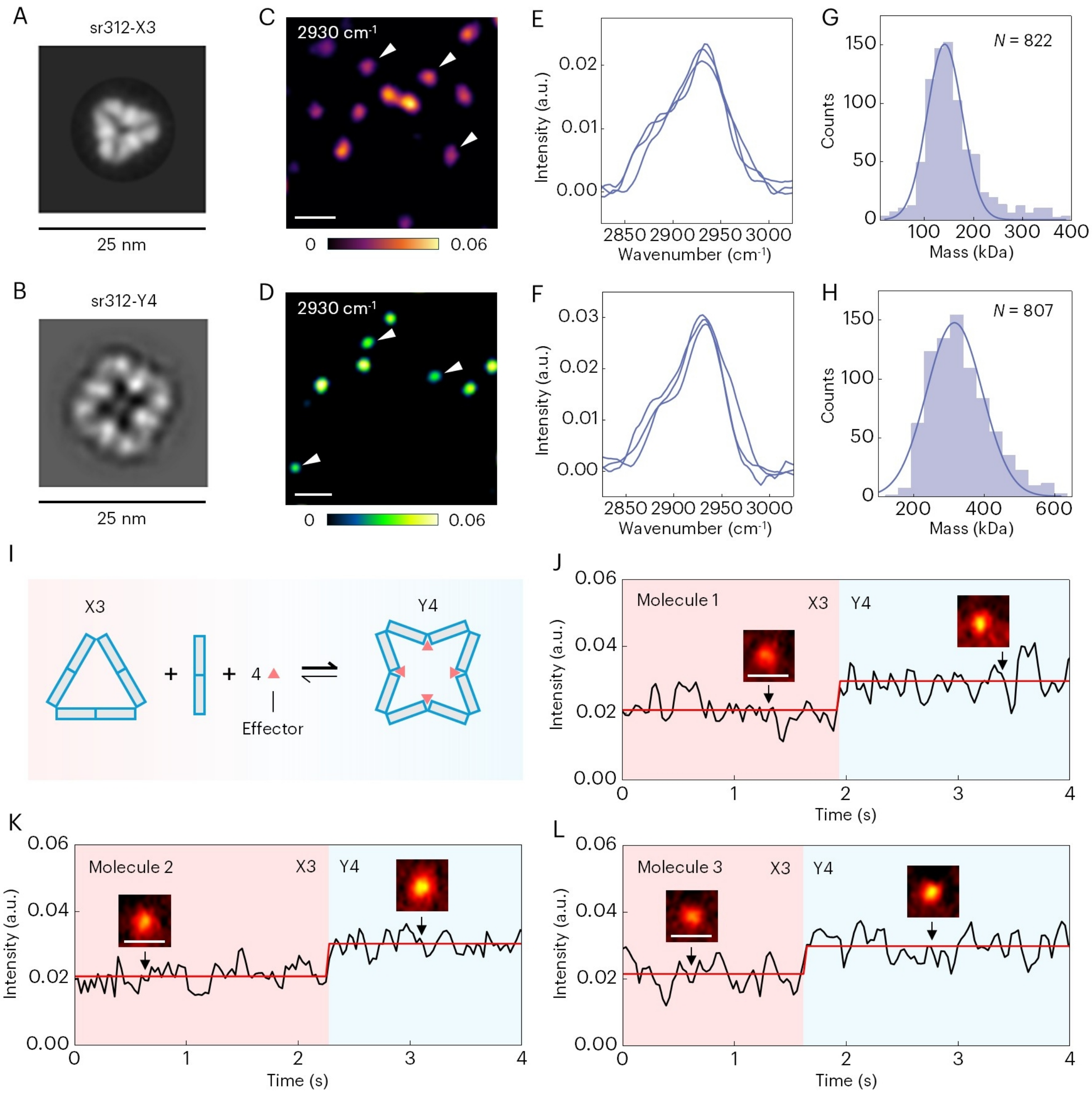

**Fig. 5. Monitoring conformational dynamics of de novo designed allosteric proteins in native environment by video-rate SRPSCAT.**
(**A** and **B**) nsEM for sr312 in the unbound X3 state (A) and peptide-bound Y4 state (B).
(**C** and **D**) SRPSCAT imaging of sr312 in X3 state (C) and Y4 state (D) at 2930 $cm^{-1}$.
(**E** and **F**) SRPSCAT spectra of single sr312 in X3 state (E) and Y4 state (F).

(**G** and **H**) SRPSCAT measurements of sr312 mass in X3 state (G) and Y4 state (H). The sequence mass of the sr312 monomer and the effector are 49.1 kDa and 31.4 kDa, respectively.
(**I**) Schematic of conformational change of sr312 from X3 state to Y4 state.
(**J** to **L**) Representative time trajectories of individual sr312 molecules monitored at 2930 $cm^{-1}$ showing conformational changes from X3 state to Y4 state in the presence of the effector. Scale bars, 1 μm.

## SRPSCAT detection of single DNA in individual viruses

Adeno-associated virus (AAV) vectors are currently one of the most promising vehicles for delivering gene therapeutics (*54*), whereas their clinical translation is frequently hindered by the structural heterogeneity inherent in viral production. Distinguishing between therapeutically active full capsids and process-related impurities, such as empty capsids or damaged vectors, is critical for ensuring potency and minimizing immune responses (*55*). However, conventional quality control methodologies often struggle to resolve this structural complexity at the single-particle level.

To address this challenge, we utilized SRPSCAT to achieve non-invasive, high-throughput and chemically specific imaging of individual AAVs. By targeting the intrinsic vibrational signatures of viral proteins and nucleic acids, SRPSCAT allows for the precise differentiation of full and empty capsids without labeling. As a demonstration, we studied AAV9 empty capsids and full capsids with a single DNA genome of 4.7 kb (fig. S17A). Through SPRSCAT hyperspectral imaging of individual empty capsids and full capsids (fig. S17, B and C), we can statistically analyze the protein and DNA component in empty and full capsids by their intensity difference at 2930 and 2959 $cm^{-1}$, respectively (fig. S17, D and E). A more significant difference was observed at 2959 $cm^{-1}$ because it represented the C–H stretching band of DNA (*56*). The minor difference shows at 2930 $cm^{-1}$ is probably due to the broad band of the C–H vibration of DNA. This can be confirmed by the single-particle spectra with peak fitting, where the DNA component is well distinguished in single full capsid (fig. S17F). These results show that SRPSCAT allows high-speed screening and chemical analysis of biomolecules and nanoparticles with single-molecule sensitivity.

## Discussion

SRPSCAT achieves quantitative bond-selective imaging of biomolecular structures and interactions at the single-molecule level in native environments by incorporating chemical specificity of SRP and single-molecule sensitivity of iSCAT. Together with AI-assisted data analysis, SRPSCAT has enabled the mass measurement of single proteins down to tens of kDa and Raman fingerprinting of protein secondary structures. Moreover, we demonstrated real-time observation of protein-protein interactions and conformational dynamics in both natural proteins and computationally designed proteins. Moving beyond the studies of protein structures and interactions, we show the broad applications in chemical imaging of nucleic acids and nanoparticles.

Compared to fluorescence-based single-molecule vibrational imaging methods such as SREF (*18*) and BonFIRE (*57*), SRPSCAT deciphers the structure and function of the target molecule in a label-free manner. Meanwhile, SRPSCAT highlights universal applicability and does not suffer from photobleaching and photoblinking. Though surface-enhanced Raman spectroscopy has allowed label-free single-molecule studies, SRPSCAT bypasses the need of plasmonic substrates and is generally applicable to quantitative biomolecule imaging under natural environments. In addition, SRPSCAT provides significant molecular fingerprint information compared to elastic scattering approaches. Although iSCAT combined with immunoaffinity assay can recognize specific bionanoparticles, it requires delicate device design and surface functionalization (*58*). Instead, SRPSCAT can directly identify different biomolecules in solution and offer Raman spectroscopy in the C–H and fingerprint region, and potentially the cell-silent region, providing abundant vibrational information. Besides in vitro study of single molecules, SRPSCAT can be extended to study the dynamic processes in surface chemistry and sensitively detect the chemical components on cell membrane or cell wall.

Further improvements would promise even better performance of SRPSCAT in several aspects. Given that the photothermal effect is sensitive to the environment, SRPSCAT under cryogenic condition (*59*, *60*) or in special medium (*61*) may lead to vibrational imaging of a single bond. These technical advances can be coupled with novel AI-denoising algorithms to help suppress the noise and thus improve sensitivity. Additionally, iSCAT detection provides phase information along axial direction near the interface. Height-sensitive 3D localization and tracking of single molecules could be achieved in combination with computational methods to reconstruct axial information (*23*, *62*). Furthermore, SRPSCAT system is compatible with microfluidics and live-cell imaging chambers, enabling high-throughput detection of multiple analytes in biological applications. Collectively, we envision that SRPSCAT microscopy will become a key tool for quantitative analysis of biomolecular structures, interactions and dynamics at the single-molecule level.

**Acknowledgments:**

**Funding:** This work was supported by NIH R35GM136223 and R01EB035429 to J.-X.C.; Breakthrough Fund Ligases Program to X.W.; the Howard Hughes Medical Institute (D.B.). AFM imaging was performed in part at the Harvard University Center for Nanoscale Systems, a member of the National Nanotechnology Coordinated Infrastructure Network, which is supported by the National Science Foundation under NSF Award No. ECCS-2025158.

**Author contributions:**

Conceptualization: P.-T.L., J.-X.C.

Methodology: P.-T.L., Y.Z., Q.X., G.D., A.P., X.W., J.A., H.L

Investigation: P.-T.L., Y.Z., Q.X., G.D., A.P.

Visualization: P.-T.L.

Supervision: L.J., D.B., J.-X.C.

Writing – original draft: P.-T.L., J.-X.C.

Writing – review & editing: All authors

**Competing interests:** Authors declare that they have no competing interests.

**Data and materials availability:** All data are presented in the manuscript or the supplementary materials.

**Supplementary Materials**

Materials and Methods

Supplementary Text

Figs. S1 to S17

References (*63–70*)

Movies S1 to S5